\documentclass{article}
\usepackage{spconf,amsmath,graphicx}
\usepackage[colorlinks = true, linkcolor=blue]{hyperref}
\usepackage[all]{hypcap}
\usepackage[printwatermark]{xwatermark}
\usepackage{xcolor}
\usepackage{graphicx}
\usepackage{lipsum}
\usepackage{fancyhdr}

\usepackage{enumitem}
\setlist{nosep, leftmargin=14pt}

\usepackage{mwe} 
\usepackage{gensymb}
\usepackage{multirow}
\usepackage{bm}

\title{Leveraging Clinically Relevant Biometric Constraints to Supervise A Deep Learning Model for the accurate Caliper Placement To obtain sonographic measurements of the fetal brain}
\name{\parbox{\linewidth}{\textbf{Hari Shankar}$^{1}$, Adithya  Narayan$^{1}$, Shefali Jain$^{2}$, Divya Singh$^{3}$, Pooja Vyas$^{4}$, Nivedita Hegde$^{5}$, Purbayan Kar$^{1}$, A Lad$^{1}$, Jens Thang$^{6}$, Jagruthi Atada$^{1}$, 
Duy Nguyen$^{6}$, PS Roopa$^{5}$, \textbf{Akhila Vasudeva}$^{5}$, \textbf{Prathima Radhakrishnan}$^{2}$, \textbf{Sripad Krishna Devalla}$^{6}$,}}
\address{$^{1}$ Origin Health India, Bengaluru, Karnataka, India \\
         $^{2}$ Bangalore Fetal Medicine Center, Bengaluru, Karnataka, India \\
         $^{3}$ Prime Imaging and Prenatal Diagnostics, Chandigarh, India \\
         $^{4}$ Jaslok Hopsital  and Research Center, Mumbai, Maharashtra, India \\
         $^{5}$ Kasturba Medical College, Manipal Academy of Higher Education, Manipal, Karnataka, India \\
         $^{6}$ Origin Health, Singapore}
\begin{document}
\pagestyle{fancy}
\fancyhf{}

%
\maketitle
\begin{abstract}
Multiple studies have demonstrated that obtaining standardized fetal brain biometry from mid-trimester ultrasonography (USG) examination is key for the reliable assessment of fetal neurodevelopment and the screening of central nervous system (CNS) anomalies. Obtaining these measurements is highly subjective, expertise-driven, and requires years of training experience, limiting quality prenatal care for all pregnant mothers. In this study, we propose a deep learning (DL) approach to compute 3 key fetal brain biometry from the 2D USG images of the transcerebellar plane (TC) through the accurate and automated caliper placement (2 per biometry) by modeling it as a landmark detection problem. We leveraged clinically relevant biometric constraints (relationship between caliper points) and domain-relevant data augmentation to improve the accuracy of a U-Net DL model (trained/tested on: 596 images, 473 subjects/143 images, 143 subjects). We performed multiple experiments demonstrating the effect of the DL backbone, data augmentation, generalizability and benchmarked against a recent state-of-the-art approach through extensive clinical validation (DL vs. 7 experienced clinicians). For all cases, the mean errors in the placement of the individual caliper points and the computed biometry were comparable to error rates among clinicians. The clinical translation of the proposed framework can assist novice users from low-resource settings in the reliable and standardized assessment of fetal brain sonograms.
 
\end{abstract}
\section{Introduction}
\label{sec:intro}
Fetal anomalies of the central nervous system (CNS) are one of the most common types (1.4-1.6 per 1000 live births), resulting in 3-6\% stillbirths, low 5-year survival rates, and lifelong physical and mental disabilities \cite{malinger_isuog_2020}. Recent advancements in obstetric ultrasonography (USG) practices and fetal medicine research have demonstrated the benefits of standardized evaluation, including reliable and consistent biometry in the timely screening of these anomalies. Specifically, the key biometry obtained from mid-trimester USG examinations of the transcerebellar plane (TC, fetal brain axial view) is critical for assessing neurodevelopment and screening posterior-fossa anomalies. Despite recommendations and guidelines from international clinical organizations, \cite{malinger_isuog_2020}, the accurate manual placement of the digital calipers in USG images to obtain these biometry (distance between the calipers) is extremely experience-driven and highly subjective. The problem further exacerbates in setting with high-volume and low patient-specialist ratios. 

Computer-aided approaches to assist the automated fetal biometry have traditionally used statistical modeling \cite{waechter-stehle_learning_2016}, and custom computer vision functions \cite{sciortino_automatic_2017} that limit their generalizability, robustness to real-world image variance, and reusability (algorithms specific to biometry). With the advent of deep learning (DL), the concept of computing biometry from automated segmentation masks became widespread, but preparing segmentation datasets is expensive and time-consuming. Further, the reliability of the biometry is contingent on the quality of automated segmentation masks. To circumvent these limitations, the concept of DL-based landmark detection \cite{noothout_cnn-based_2018} to directly obtain the key/caliper points to compute the measurements/biometry has become popular.

In the context of DL based landmark detection in medical imaging, studies have investigated the use of  patch-based methods \cite{noothout_cnn-based_2018}, multi-task learning \cite{xu_less_2018}, cascaded networks (coarse to fine detection) \cite{bier_x-ray-transform_2018}, and attentive pyramid fusion modules \cite{chen_cephalometric_2019}. However, they don't consider the relationship between the landmark points and hence lack a global context essential for computing biometry. To model the relationship between the landmark points, Zhang \textit{et al.} \cite{zhang_context-guided_2020} and Tuysuzoglu \textit{et al.} \cite{tuysuzoglu_deep_2018} segmented the associated contours and regions along with landmark detection and demonstrated the effectiveness of this idea. Recently,  Wei Liu \textit{et al.} \cite{liu_landmarks_2020} proposed an implicit relation loss by regressing the distance vector between related landmarks to improve the accuracy.

In this study, we developed an end-to-end deep learning (DL) approach for the automated caliper placement of 6 points in the TC plane of the fetal brain to compute 3 key biometry (transcerebellar diameter [TCD]; cisterna magna size [CMS]; nuchal fold thickness [NFT]). We formulate the task as a landmark detection problem to predict the caliper points. We propose the 
concept of biometric constraint-based supervision (BCS). This clinically intuitive and relevant method explicitly models the relationship between the landmarks by segmenting the lines connecting the related landmark points (required for computing the biometry). Through multiple experiments, we analyze the effect of the DL backbone on the performance, benefits of the proposed BCS, and impact of domain-relevant data augmentation. We clinically validate by benchmarking against 7 experienced (\(>\)5 years) clinicians routinely performing fetal USG. Lastly, we also demonstrate the generalizability by repeating the entire process (training and testing) to predict 2 caliper points in another axial plane (transventricular [TV]) to compute the atrial width (AW) of the lateral ventricle (LV).

\begin{figure*} [ht!]
\begin{center}
\begin{tabular}{c} 
\includegraphics[width=0.9\textwidth]{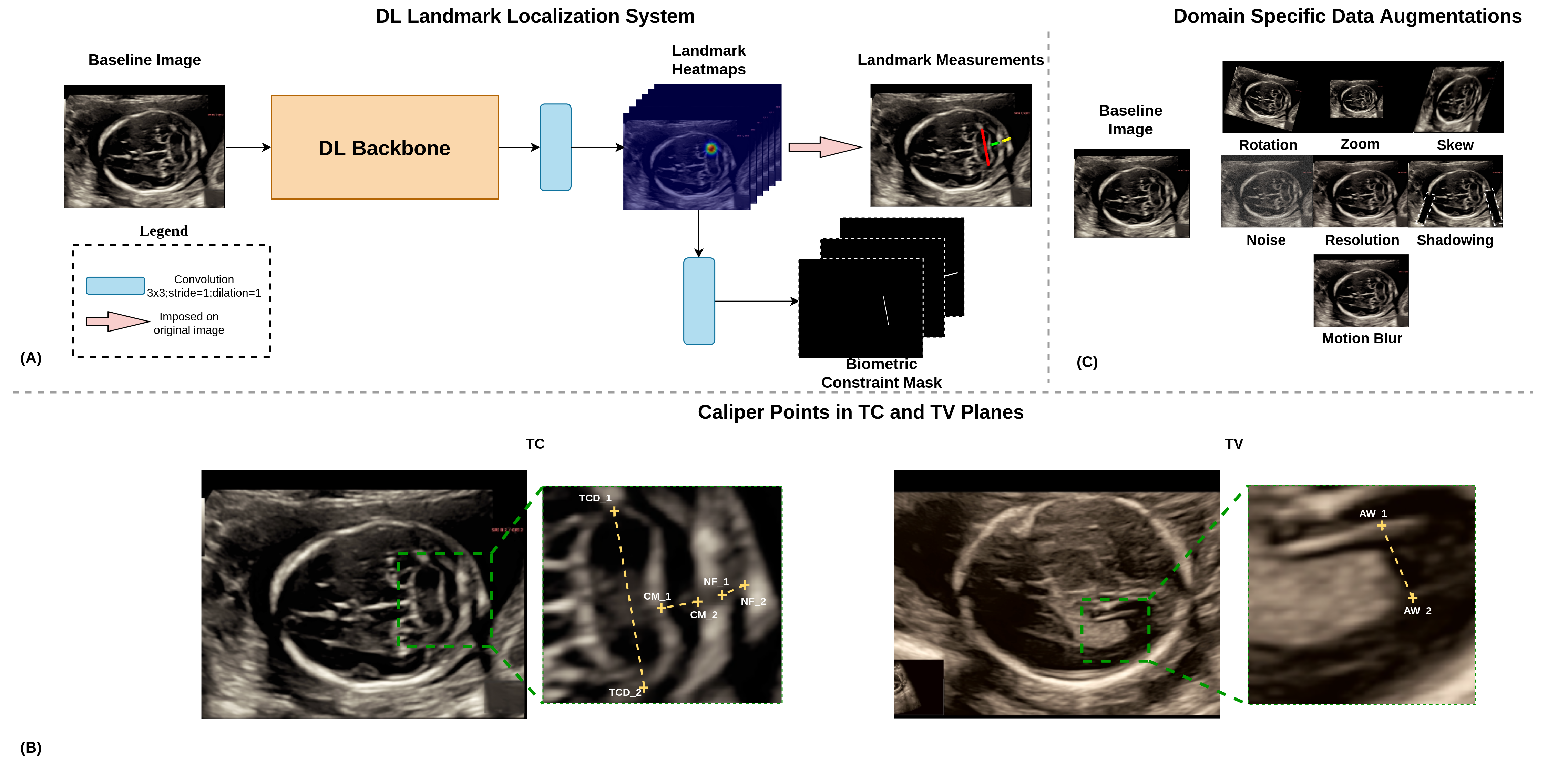}
\end{tabular}
\end{center}
\caption{ \label{fig1} 
(A) The U-Net inspired DL framework used in this study is shown. The landmark measurement figure's red, green, and yellow lines represent TCD, CMS, and NFT, respectively. (B) The caliper points in TC and TV planes are shown. The yellow dotted lines represent the biometry, and the plus symbols at the end of lines represent the caliper points required for the biometry. (C) Domain-specific data augmentation (1: baseline; 2: rotation; 3: zoom; 4: skew; 5: speckle noise; 6: resolution; 7: shadowing [highlighted in white box]; 8: motion blur) are shown.}
\end{figure*}

\section{DATASET}
\label{sec:format}

\subsection{Patient recruitment}
\label{sec:patientrec}

We used two datasets of the images of the TV plane (TV dataset) and TC plane (TC dataset) in this study. The TC dataset was used for all experiments, while the TV dataset was used to demonstrate the generalizability of the proposed approach. A total of 1192 images (596 TC, 596 TV) were retrospectively obtained from a total of 473 mid-trimester USG examinations (18-24 weeks; transabdominal scans) at 3 centers (2 tertiary referral centers and 1 routine imaging center) using GE Voluson E8, S10, and P8 (General Electric Healthcare, Chicago, Illinois, USA) USG machines. All images were taken and reviewed by fetal medicine specialists for quality and correctness. Additionally, only live singleton fetuses that did not exhibit any growth anomalies were included in this study. Due to the retrospective nature of the study, as per the tenets of the Declaration of Helsinki, all informed consent was waived and data were completely anonymized after approval from respective ethics committees.

\subsection{Dataset preparation}
\label{sec:dataprep}
For both TV and TC datasets, 453 images were split into training/validation (88:12) sets to train and fine-tune the DL model, with no patient overlap between the splits. The ground truth caliper points were prepared for the training and validation datasets by medical expert annotators based on international guidelines \cite{malinger_isuog_2020} and reviewed by 2 independent fetal medicine specialists. In the TC plane \hyperref[fig1]{Fig. 1(B)}, 3 biometrics (2 caliper points each; 6 points overall) including \textbf{(1)} TCD (maximal diameter between the cerebellar hemispheres); \textbf{(2)} NFT (distance from the outer edge of the occipital bone to the outer edge of the skin along the midline); and \textbf{(3)} CMS (anterioposterior diameter of the cisterna magna) were annotated. Similarly, in the  TV plane \hyperref[fig1]{Fig. 1(B)}, 1 biometry (2 caliper points) were annotated:  \textbf{(1)}  AW (lateral ventricle’s width taken at the glomus of the choroid plexus [inner edge to inner edge] perpendicular to the ventricular cavity axis).

For testing, we used a fully-independent test set of 143 images per dataset (143 pregnancies) obtained from 3 machines at 3 centers. The caliper positions of the respective biometry for each image in the test set were manually annotated by 7 experienced clinicians. Furthermore, to account for intra-operator variability, each caliper position was annotated twice (3 days apart and all data randomized to minimize perceptual biases), and the mean was taken as the ground truth.

All images were resized to 320 (height) x 576 (width) pixels. No additional pre-processing was performed to preserve the inherent variability (i.e., gain, zoom, contrast, speckle noise, patient-specific probe settings, etc.) in the dataset.

\section{METHOD}

\label{sec:method}

\subsection{Deep learning method description}
\label{sec:deep learning architecture}
We propose a fully convolutional end-to-end DL approach \hyperref[fig1]{Fig. 1(A)} to simultaneously place 6 caliper points for obtaining 3 key biometry. The framework consisted of two-outputs \hyperref[fig1]{Fig. 1(A)}: \textbf{(1)} landmark heatmaps and \textbf{(2)} biometric constraint masks. We converted each landmark point to a normalized heatmap of a 2D Gaussian distribution (with the standard deviation $\sigma$ as a tunable hyperparameter) centered around the landmark pixel to obtain the 6 channel (1 per caliper point) landmark heatmaps. The pair of landmark points corresponding to single biometry were connected by a line (with width w as tunable hyperparameter; biometric constraint supervision [BCS]) to get the 3 channel biometric constraint mask corresponding to the 3 biometry as shown in \hyperref[fig1]{Fig. 1(A)}.

A 2D USG image of size 320 x 576 was first passed through the backbone network to extract the high-level features. The landmark heatmaps and the biometric constraint masks were subsequently regressed from the high-level features through one convolutional layer each. The loss for the system was then computed as a weighted sum of the loss from the landmark heatmap (\(L_{H}\)) and the BCS loss (\(L_{BCS}\)).
\begin{equation}
    L = L_{H} + \alpha L_{BCS}
\end{equation}

where \(\alpha\) is the loss weight. Both \(L_{H}\) and \(L_{BCS}\) were evaluated as the cross-entropy loss between predicted and the ground truth heatmaps/masks.

During inference, each predicted landmark channel \(h_{i}\) was normalized and thresholded at 0.85 (empirically found), and all the connected components were obtained. The confidence score of a connected component, denoted by a binary mask \(m \in {\{0, 1\}}^{w*h}\) was taken as the sum of the associated values in the predicted channel, i.e. \(conf(m) = sum(m \odot h_{i})\). The final coordinates were calculated as the barycenter of the most confident connected component.

\subsection{Domain-specific data augmentation}
\label{sec:DA}
The quality of fetal USG images are often affected by 2 major sources of variance \cite{feldman_us_2009}: \textbf{(1)} imaging based (e.g., acoustic shadows, poor structure differentiability [contrast], speckle noise) and \textbf{(2)} operator based (e.g., incorrect zoom, centering, poor-resolution). To counter these effects and improve the generalizability of the DL model , extensive domain-specific online data augmentation (DA) were performed. These included multiplicative Gaussian noise ($\mu = 0, \sigma=0.1$) to simulate speckle noise,  repeated downsampling/upsampling ($0.3-0.7$) to simulate quality reduction , acoustic shadowing (extremities of the cranium; 2 patches of size: 75-100 pixels, angle: 15-20 degrees), motion blur (kernel size: 50x50 pixels), zoom ($0.6-1.0$), and affine (rotation: $\pm 20\degree$, translation: $40-60$ pixels across height/width) and shear transformations ($0-0.3$) to account for operator variability.

\begin{table*}[ht!]
\centering
\caption{Quantitative comparison of the mean absolute error in millimeters to analyze the effect of the proposed DA, BCS across three different backbones and benchmarking the BCS method against Wei Liu \textit{et al.} \cite{liu_landmarks_2020}. TCD\_1, TCD\_2 indicates the first and second caliper points required to measure TCD. Similar notation is used for CMS and NFT.}. 
\scalebox{0.75}{
\begin{tabular}{|l|c|c|c|c|c|c|c|}
\hline
Methods & TCD\_1 & TCD\_2 & CMS\_1 & CMS\_2 & NFT\_1 & NFT\_2 & Mean\\ 
\hline
\textbf{UNet + DA + BCS}         & \bm{$1.24\pm0.65$}          & \bm{$1.56\pm0.68$}        & \bm{$1.34\pm0.72$}    & \bm{$1.87\pm1.04$}          & \bm{$2.81\pm1.05$}        & \bm{$3.10\pm1.20$}    & \bm{$1.98\pm0.89$} \\
\hline 
\textbf{Wei Liu \textit{et al.} }\cite{liu_landmarks_2020}
& $1.44 \pm 0.83$               & $1.58 \pm 0.79$                  & $1.60\pm 0.81$  
& $1.94 \pm 1.07$               & $3.05 \pm 0.98$& $3.08 \pm 1.10$   & $2.11 \pm 0.93$ \\
\hline
UNet + BCS
& $1.25\pm0.70$               & $1.80\pm0.80$                  & $1.52\pm0.75$  
& $2.20\pm1.21$               & $3.41\pm1.39$& $3.54\pm1.48$   & $2.28\pm1.05$ \\
\hline
UNet + DA
& $1.35\pm0.74$               & $1.59\pm0.62$                  & $1.47\pm0.77$  
& $1.99\pm1.10$               & $3.04\pm1.05$                  & $3.19\pm1.16$ 
& $2.10\pm0.90$ \\
\hline
HR-Net + DA + BCS
& $1.18\pm0.62$               & $1.70\pm0.71$                  & $1.42\pm0.92$  
& $1.92\pm1.14$               & $2.87 \pm1.02$                  & $3.09 \pm 1.03$ 
& $2.03 \pm 0.90$ \\
\hline
HR-Net + BCS
& $1.33 \pm 0.75$               & $1.66 \pm 0.74$                  & $1.57 \pm 0.98$
& $2.11 \pm 1.25$               & $3.45 \pm 1.41$                  & $3.46 \pm 1.48$ 
& $2.26 \pm 1.10$ \\
\hline
HR-Net + DA
& $1.33 \pm 0.68$               & $1.49 \pm 0.60$                  & $1.49 \pm 0.78 $  
& $2.02 \pm 1.06$               & $3.21 \pm 1.11$                  & $3.64 \pm 1.25$ 
& $2.19 \pm 0.91$ \\ 
\hline
Hglass + DA + BCS
& $1.22 \pm 0.68$               & $1.56 \pm 0.71$                  & $1.47 \pm 0.84$  
& $2.06 \pm 1.15$               & $3.05 \pm 1.01$                  & $3.22 \pm 1.06$ 
& $2.09 \pm 0.90$ \\
\hline
Hglass + BCS
& $1.43 \pm 0.85$               & $1.79 \pm 0.96$                  & $1.94 \pm 1.28$  
& $2.65 \pm 1.59$               & $3.35 \pm 1.48$                  & $3.48 \pm 1.42$ 
& $2.44 \pm 1.26$ \\
\hline
Hglass + DA
& $1.26 \pm 0.67$               & $1.61 \pm 0.70$                  & $1.63 \pm 0.84$  
& $2.21 \pm 1.18$               & $3.14 \pm 1.13$                  & $3.29 \pm 1.21$ 
& $2.19 \pm 0.95$ \\
\hline
\end{tabular}
}
\label{table1}
\end{table*}
\begin{figure*} [ht!]
\begin{center}
\begin{tabular}{c} 
\includegraphics[width=0.9\textwidth]{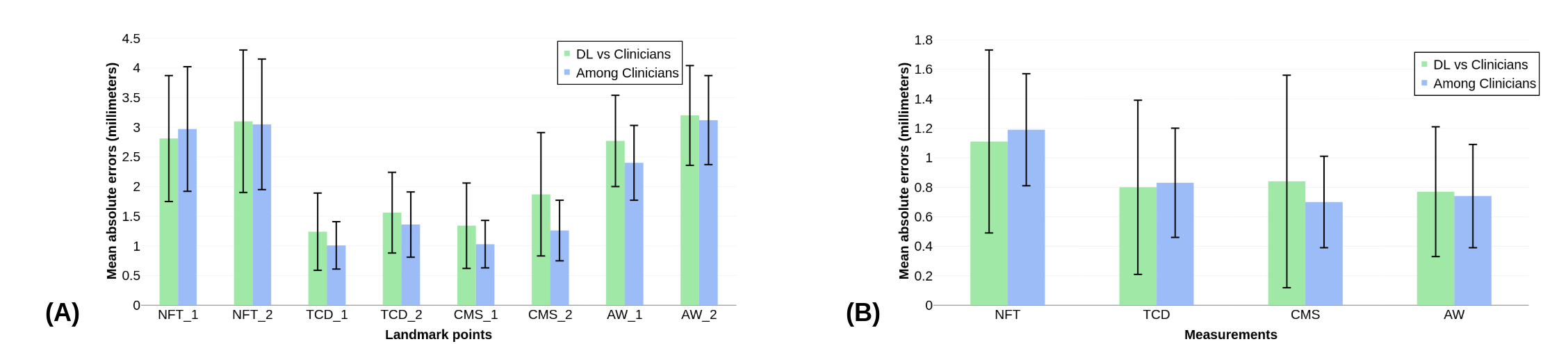}
\end{tabular}
\end{center}
\caption{ \label{fig2} 
Comparison of the mean absolute errors (millimeters) of the proposed DL method with the clinicians and the inter-rater errors within the clinicians is shown. (A) - Errors in the caliper positions; (B) - Errors in the actual biometric measurements obtained from the caliper positions.}
\end{figure*}

\section{Experiments and Results}
\label{sec:experiemtns}
The proposed framework was implemented using Pytorch 1.7.0 and was trained/tested on an NVIDIA Tesla T4 with CUDA v11.0 and cuDNNv7.6.5. The framework was trained end-to-end with Adam optimizer with a fixed learning rate of 0.0001 for 150 epochs for all the experiments. The best standard deviation of heatmaps \(\sigma\)(2),  loss weight \(\alpha\)(1e-3), and the line width w(6) were found empirically. All the experiments for the TC plane dataset were performed across three different backbone networks namely U-Net \cite{ronneberger_u-net_2015}, Stacked Hourglass \cite{newell_stacked_2016}, HRNet \cite{wang_deep_2020} to study the performance across backbones. The mean absolute error between the predicted caliper positions and the ground truth from each of the 7 clinicians were calculated and used as the metric to compare different experiments. We performed the following experiments:
\begin{itemize}
\item \textbf{Experiment 1:} Ablations were performed on all the 3 backbones to assess the effect of DA and BCS. Also, to analyze the effectiveness of BCS in modeling the relationship between landmarks, the proposed method was benchmarked against the recent work by Wei Liu \textit{et al.} \cite{liu_landmarks_2020} which proposes a relation loss to model the relationship between landmarks by regressing the distance between related landmarks (Refer \hyperref[table1]{Table 1}, Column 1).
\item \textbf{Experiment 2:} Ablations and benchmarking in Experiment 1 were performed on the best backbone for the TV dataset to analyze the approach generalizability. 
\end{itemize}

From \hyperref[table1]{Table 1}, we can infer that, in general, the use of the proposed domain-specific DA and BCS offered a 12\% and 4\% average improvement across three different DL backbones. Specifically, the U-Net model performed the best among the three backbone networks. The proposed BCS offered a 6\% improvement over Wei Liu \textit{et al.} \cite{liu_landmarks_2020} method of regressing the distance between related landmarks. The mean caliper placement errors (in millimeters) for the \textbf{U-Net + DA + BCS}, U-Net + BCS, U-Net + DA on the TV dataset were\textbf{ \(2.985 \pm 0.80\)},  \(3.075 \pm 0.93\) and \(3.175 \pm 0.94 \) respectively indicating an improvement of 3\% and 6\% with the addition of DA and BCS respectively. The proposed method outperforms the benchmarking method by 5\% in the TV dataset as well.

From \hyperref[fig2]{Fig. 2}, it can be inferred that the mean absolute errors of the DL model are very comparable to the errors among the clinicians for all the caliper positions. Further, the 3 key biometry of the TC plane and the AW measurement of the TV plane were calculated from the landmark points, and the mean absolute errors of the measurements were also computed. The mean errors among the clinicians and the mean errors of the DL model across all 4 measurements were \(0.86 \pm 0.35\) mm and  \(0.88 \pm 0.59\) mm, respectively. The intra-class correlation coefficients (ICC) values of the DL measurements with the clinicians' measurements were calculated based on two-way random, absolute agreement, and average rater policy. The best DL (U-Net + DA +BCS) model achieved an ICC score of 0.98 for TCD, 0.79 for NF, 0.83, for CMS and 0.89 for AW. The high ICC values indicate good to excellent reliability for all 4 measurements when benchmarked with 7 clinicians.

\section{Conclusion}
We presented a custom DL-based approach for the accurate caliper placement to compute 3 key biometry (TCD, CMS, NFT). We also demonstrated the effectiveness of our proposed BCS and DA through extensive experiments across three different backbones and two different datasets (TC and TV planes). When tested on 143 images from an unseen test set, the model offered a mean caliper placement error of $1.98 \pm 0.89$ mm against 7 experienced clinicians. We believe that the successful clinical translation of the proposed framework can assist novice users in the accurate and standardized assessment of fetal brain USG examinations to aid the screening of CNS anomalies. The limitations of this study include limited dataset size and device variability (only 3 USG devices). The robustness of the proposed approach across different gestation ages and anomaly cases was not studied. The problem of identifying the standard TV and TC planes will be studied in future works to ease the clinical translation of the proposed framework.
 
\section{Author Information}

Akhila Vasudeva, Prathima Radhakrishnan, and Sripad Krishna Devalla contributed equally as clinical, clinical, and technical Principal Investigators, respectively. All three are corresponding authors.

\bibliographystyle{IEEEbib}
\bibliography{strings,refs}

\end{document}